\documentclass[10pt]{iopart}

\usepackage{amssymb}
\usepackage{graphicx}

\begin{document}
\title[The effect of thermophoresis on the discharge parameters\ldots]{The effect of thermophoresis on the discharge parameters in complex plasma experiments.}

\author{Victor Land, Jorge Carmona-Reyes, James Creel, Jimmy Schmoke, Mike Cook, Lorin Matthews, and Truell Hyde}

\address{Center for Astrophysics, Space Physics, and Engineering Research, Baylor University, Waco, TX, USA 76798-7316}
\ead{victor\_land@baylor.edu}

\begin{abstract}
Thermophoresis is a tool often applied in complex plasma experiments. One of the usual stated benefits over other experimental tools is that changes induced by thermophoresis neither directly depend on, nor directly influence, the plasma parameters. From electronic data, plasma emission profiles in the sheath, and Langmuir probe data in the plasma bulk, we conclude that this assumption does not hold. An important effect on the levitation of dust particles in argon plasma is observed as well. The reason behind the changes in plasma parameters seems to be the change in neutral atom density accompanying the increased gas temperature while running at constant pressure.
\end{abstract}


\maketitle

\section{Thermophoresis in complex plasma}

Partially ionized gases containing small solid particles, called complex plasmas, are a widespread phenomenon in space, found for instance in molecular clouds, planetary atmospheres, planetary rings, and comets \cite{Mendis1994,Merlino2004}. In the laboratory, complex plasma has become a promising tool to study the physics of solid state systems, fluid dynamics and turbulence, even biophysics, on a level accessible with ordinary optical techniques \cite{Melzer1996, Morfill2004, Tsytovich2007}.

Unless experiments are performed during parabolic flights, or under micro-gravity conditions on the International Space Station \cite{Morfill1999}, gravity dominates the force equilibrium in complex plasma, resulting in the formation of two-dimensional crystal structures in places where the electrostatic force acting on the particles balances gravity. An alternative is to extend the confinement of the particles vertically by adding a dielectric (glass) box in the discharge \cite{Arp2004}. In these experiments, thermophoresis played an important role in reducing the effect of gravity, and 3-dimensional crystals were produced.

The thermophoretic force depends on the establishment of a temperature gradient in the background gas, resulting in a net imbalance of the momentum imparted by neutral atoms colliding with a particle in the gas. Therefore, it has to date been regarded as a tool independent of the plasma parameters, which in complex plasma experiments are hard to measure, since they can become strongly coupled to the presence of particles in the plasma \cite{Stefanovic2003}. As a consequence, changes in discharge characteristics (like the DC bias, or absorbed power) accompanying temperature changes are usually considered insignificant and are not specified in complex plasma experiments involving thermophoresis, see for instance \cite{Rothermel2002}.

This paper reports on complex plasma experiments involving thermophoresis performed in a modified Gaseous Electronic Conference (GEC) reference cell \cite{Hargis1994} at the Center for Astrophyiscs, Space Phyiscs, and Engineering Research (CASPER). Two direct observations connected with thermophoresis are a change in the DC bias on the powered electrode with increasing electrode temperature, which was briefly mentioned before \cite{Land2010}, accompanied by a change in levitation height of the particles. From electronic data, optical emission data, and Langmuir probe data, we show that the plasma properties indeed change. Using a self-consistent fluid model \cite{Land2009}, we show that the volume averaged gas-temperature increases with electrode temperature, resulting in a decrease of the neutral atom density.

\section{The complex plasma experiment at CASPER}

The experiments discussed here were performed using argon gas at low pressures (a few tens of Pascals). The upper electrode in the cell, as well as the walls surrounding the chamber, are grounded, while a 13.56 MHz radio-frequency (RF) potential is applied to the lower electrode with an amplitude of a few tens of Volts.

Once plasma is formed, spherical micron-sized melamine-formaldehyde (MF) particles are introduced from shakers situated at the top of the chamber. The particles fall through the grounded electrode, which consists of a hollow cylinder, rather than a solid electrode. At equilibrium, the particles are vertically suspended in the sheath above the bottom electrode, while horizontally confined by a cutout plate containing a cylindrical depression creating a radial electric field placed on top of the bottom electrode.

\begin{figure}[htbp]
\center
\fbox{\includegraphics[width=0.75\textwidth]{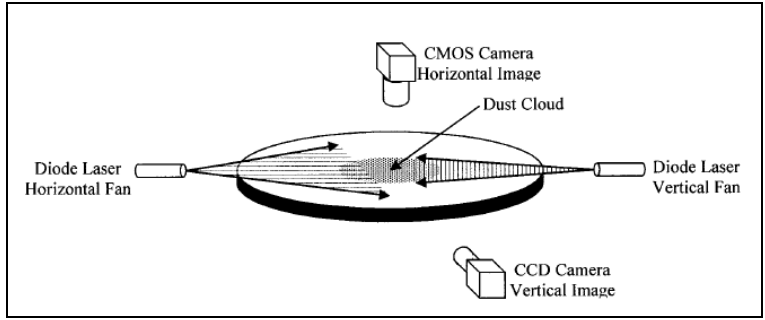}}
\caption{\label{fig:setup} A sketch of the experimental setup used. Shown are the laser/camera pairs, the lower electrode with the cutout milled in the cover plate and a cloud of dust particles suspended above the cutout.}
\end{figure}

Two diode lasers are used to illuminate the particles. The lasers are equipped with cylindrical lenses, so that thin ($\pm$ 100 micron), but wide (fan angle $>$ 5 ${}^{\circ}$) laser sheets are created. A vertical laser sheet in combination with a side-view camera allows side-view pictures of the suspended particles, whereas a horizontal laser sheet together with a top-view camera focused through the hollow upper cylinder, allows for top-view images. A schematic drawing of the experimental setup is shown in figure \ref{fig:setup}.

The lower electrode temperature is regulated by coolant liquid running through tubes which are placed inside of bores in the electrode. The temperature of the liquid is maintained by a heater/chiller with a thermostat, and is adjusted from 0${}^{\circ}$ C to +70${}^{\circ}$ in steps of 10${}^{\circ}$ C. The lower electrode temperature is measured with a thermistor attached to the bottom of the electrode and logged with a digital datalogger.

Electronic data, including the driving potential, the electrode current, and the signals from two derivative probes, measuring the derivatives of the potential and electrode current, are collected on two oscilloscopes. The latter signals are used to determine the phase angle between the potential and the current. From this angle and the first two signals, the power input into the plasma is calculated. From the electronic data, the floating DC bias on the lower electrode is obtained too, as the shift in the mean value of the driving potential data.

\section{Data obtained during thermophoresis complex plasma experiments}

This section presents the data obtained during experiments at a pressure of 200 mTorr (26 Pa). In these experiments 8.89 micron diameter MF particles were used. Initially, the DC bias was allowed to float. Upon observing the change in DC bias with electrode temperature, the experiments were repeated using an external DC power supply to fix the DC bias. We first present the measured DC bias and the levitation height of the particles above the lower electrode obtained with the side-view camera, for the two sets of experiments. We then present the additional data obtained from the data-loggers, oscilloscopes, and Langmuir probe, as well as optical emission profiles in the sheath.

\subsection{DC bias and dust levitation}

\begin{figure}[htbp]
\center
\includegraphics[width=1.0\textwidth]{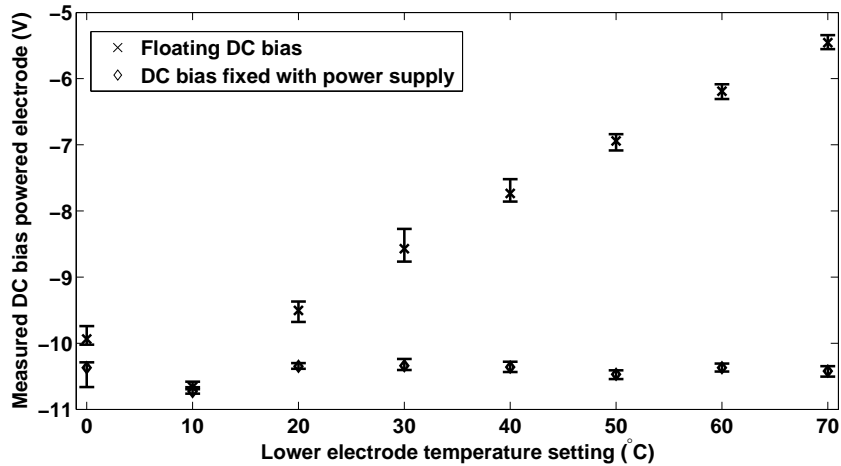}
\caption{\label{fig:DCBIAS} The mean floating DC bias measured for thermophoresis experiments with different thermostat settings. The bars indicate the range of bias potential measured during the 150 second measurement period, for each lower electrode temperature setting. $X$ refers to the floating DC bias experiment, while the $\Diamond$ refer to the experiment with the DC bias \emph{held constant} by the external DC power supply.}
\end{figure}

The measured DC bias on the lower electrode is shown for both sets of data in figure \ref{fig:DCBIAS}. When the DC bias is allowed to float, it becomes less negative with increasing lower electrode temperature. Because our discharge chamber is fairly symmetric, the induced DC bias is small; even so, the change in DC bias with the change in lower electrode temperature is significant. Even when the power supply was used, the DC bias had to be manually adjusted for each electrode temperature setting, showing that it still changed despite the connection to the power supply. Overall, this resulted in a slight variation of the DC bias, although the variation was much smaller than in the experiments with floating DC bias.

\begin{figure}[htbp]
\center
\includegraphics[width=0.9\textwidth]{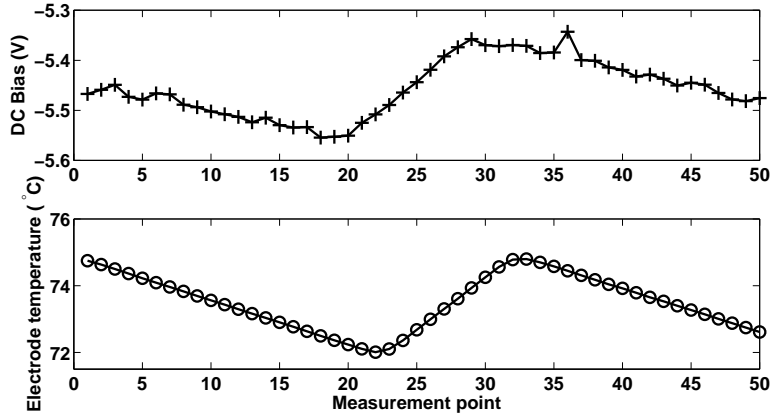}
\caption{\label{fig:changingBIAS} The electrode temperature and DC bias measured over approximately 150 seconds for one of the thermophoresis experiments described in the text. The DC bias clearly changes with electrode temperature over shorter time-scales, illustrating that the change in DC bias is not a long-term charging effect.}
\end{figure}

For each temperature setting in each experiment, the DC bias was logged for 150 seconds. The error bars in figure \ref{fig:DCBIAS} indicate the range over which the DC bias varied during these 150 seconds, whereas the symbol shows the mean. As can be seen, the DC bias varies over a rather large range. This is better illustrated in figure \ref{fig:changingBIAS}, where fifty consecutive data points are shown. The lower panel shows the registered lower electrode temperature with the thermostat set at 70${}^{\circ}$ C, while the upper panel shows the DC bias on the lower electrode. In this case, the DC bias was floating with values fluctuating around -5.45 V. Clearly, the DC bias directly follows the temperature on the powered electrode. The variations in temperature shown are due to hysteresis in the heater/thermostat system. As can be seen, there is an offset between the temperature setting on the thermostat and the mean temperature measured with the thermistor (about 4${}^{\circ}$ C at 70${}^{\circ}$ C), as well as the tendency to overshoot the set temperature (plus offset).

\begin{figure}[htbp]
\begin{minipage}[b]{0.5\linewidth}
\setlength\fboxsep{0pt}
\setlength\fboxrule{0.75pt}
\fbox{\includegraphics[scale=0.3]{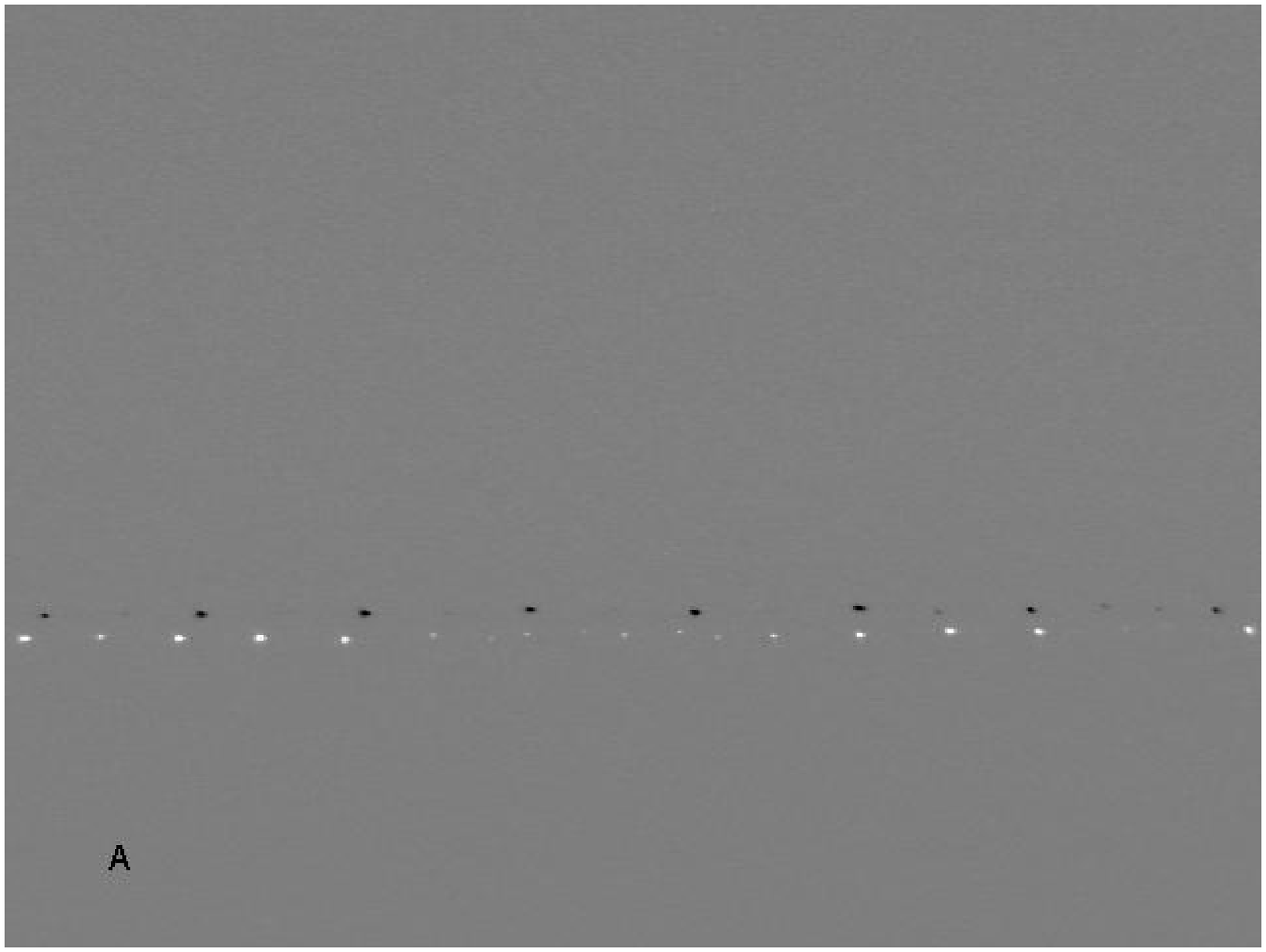}}
\end{minipage}
\hspace{1.0cm}
\begin{minipage}[b]{0.5\linewidth}
\includegraphics[scale=0.39]{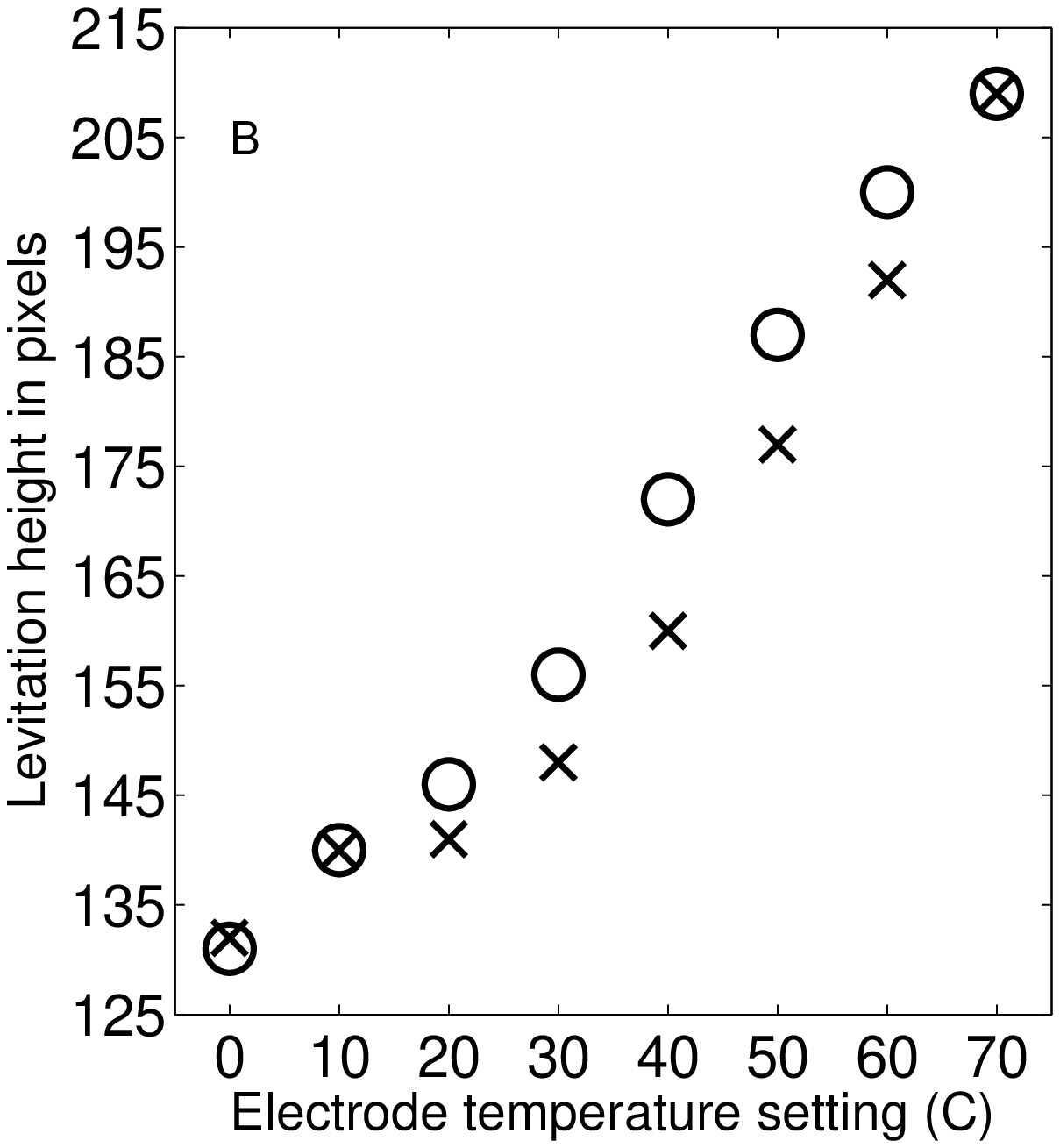}
\end{minipage}
\caption{\label{fig:dustlevitation} \textbf{A}. This image shows the result when a side-view image of dust particles levitated with floating DC bias was subtracted from an image of dust particles levitated with fixed DC bias, with the electrode held at 40${}^{\circ}$ C. The black dots correspond to the dust particles with floating DC bias, the bright dots with fixed DC bias. Clearly, a difference in levitation height is observed. \textbf{B}. The mean levitation height for experiments with fixed DC bias ($X$) and the experiments with floating DC bias ($\circ$). The size of the symbols is an indication of the spread in the mean levitation height obtained from tens of images taken during each experiment.}
\end{figure}

The levitation height of the particles obtained in both sets of experiments is shown in figure \ref{fig:dustlevitation} B, while in figure A the overlayed side-view images for both experiments at an electrode temperature of 40${}^{\circ}$ is shown. The white dots represent the particles in the experiment with fixed DC bias, the black dots the particles in the experiment with floating DC bias. There is a clear difference in levitation height of about 20 pixels. The sideview camera resolution is roughly 25 $\mu$m per pixel, so this difference in levitation height corresponds to 500 $\mu$m.

At the lowest electrode temperature setting (0${}^{\circ}$ C) the dust levitation height is the same for both sets of experiments. For increasing temperature the difference increases, coinciding with the increasing difference in the DC bias between the two sets of experiments. At the highest electrode temperature setting (70 ${}^{\circ}$), the difference in levitation height decreases again.

\subsection{Electronic data from oscilloscopes}

\begin{figure}[htbp]
\center
\includegraphics[width=0.99\textwidth]{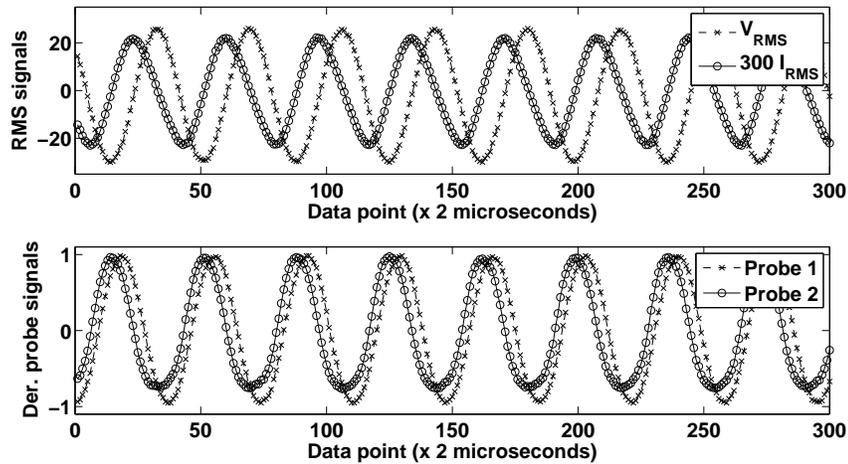}
\caption{\label{fig:E-data} Electronic data obtained from the oscilloscopes. The top panel shows the RMS potential on the lower electrode, as well as the net current to the lower electrode ($\times$ 300). The bottom panel shows the normalized derivative probe signals, which are used to obtain the phase angle $\Phi$ between the current and the potential.}
\end{figure}

To investigate whether the changes shown are related to changes in the plasma parameters, we logged the root-mean-square driving potential ($V_{RMS}$) and current ($I_{RMS}$), as well as signals from two derivative probes. The raw signals are shown in figure \ref{fig:E-data}. Each RF cycle is sampled by 37 points, giving a resolution of roughly 2 nanoseconds. The top panel shows $V_{RMS}$ and 300 $\times I_{RMS}$ (which is typically on the order of 10 mA to 100 mA). The bottom panel shows the normalized derivative probe signals, which are used to determine the phase angle $\Phi$ between $V_{RMS}$ and $I_{RMS}$. 

\begin{figure}[htbp]
\center
\includegraphics[width=0.99\textwidth]{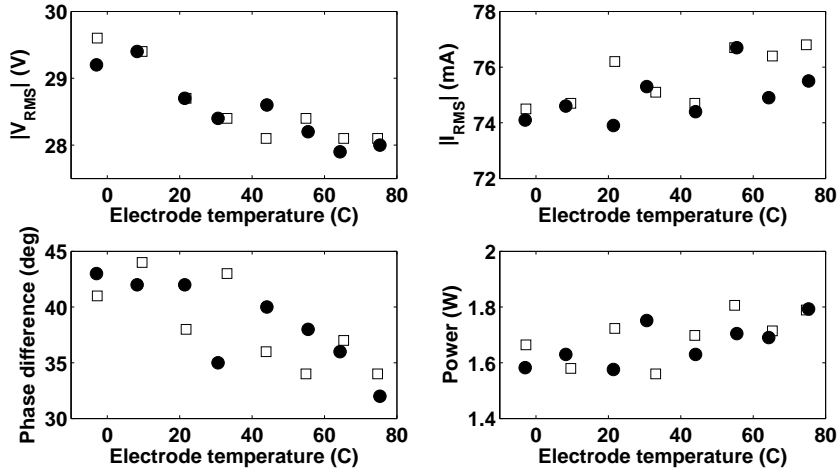}
\caption{\label{fig:E-data2} The amplitude of the RMS voltage, current, the phase angle between them and the RMS aborbed power in the plasma, versus electrode temperature for the experiments with floating DC bias ($\square$), and with fixed DC bias ($\bullet$).}
\end{figure}

Using this data, we can calculate the total power absorbed in the plasma, as $P_{RMS} = V_{RMS}I_{RMS}\cos(\Phi)$. The amplitude of the RMS voltage and current, the phase angle $\Phi$, as well as the RMS plasma power derived for both sets of experiments are shown in figure \ref{fig:E-data2}. From the results, we see that the amplitude of the RMS potential decreases, the amplitude of the RMS current increases and $\Phi$ decreases, as the electrode temperature is increased. The final result is an increase in the RMS plasma power, with a linear fit to both sets yielding an increase in power of 2 mW/${}^{\circ}$C, so that over the temperature range considered, the power increases by 10\%. 

\subsection{Bulk Langmuir probe data}
\begin{figure}[htbp]
\center
\includegraphics[width=0.8\textwidth]{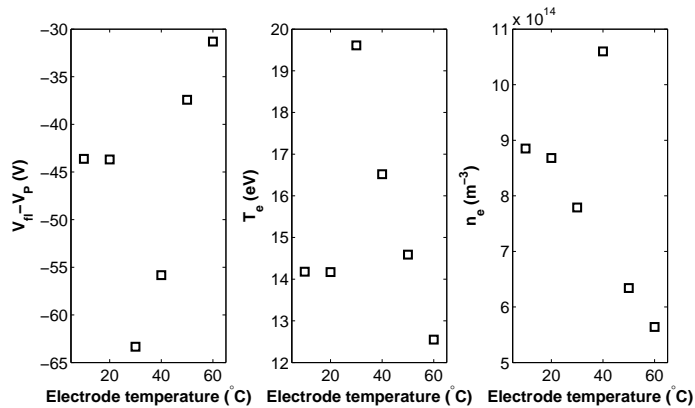}
\caption{\label{fig:Langmuirdata} The Langmuir probe data obtained in the bulk plasma. From left to right, the local floating potential, the electron temperature and the electron density are shown.}
\end{figure}

Using a SmartProbe Langmuir probe from Scientific Systems \cite{Scisys} the plasma parameters in the bulk plasma were measured while varying the electrode temperature. Due to technical constraints with the probe, the electrode temperature setting was limited between 10${}^{\circ}$ C and 60${}^{\circ}$ C. Figure \ref{fig:Langmuirdata} shows the local floating potential (defined as the floating potential with respect to ground measured by the probe minus the local plasma potential with respect to ground as measured by the probe), the local electron temperature in electron volts, and the local electron density.

A large scatter in the data can be seen, which is often the case with Langmuir probe measurements. No clear trend with the electrode temperature is apparent, although one could argue that the electron density appears to decrease with increasing electrode temperature. The electron temperature does seem to be high, since in typical plasma discharges in the GEC cell at similar discharge settings, the electron temperature is normally on the order of 1 eV \cite{Anderson1995}.

\subsection{Sheath emission profiles}

\begin{figure}[htbp]
\center
\includegraphics[width=0.7\textwidth]{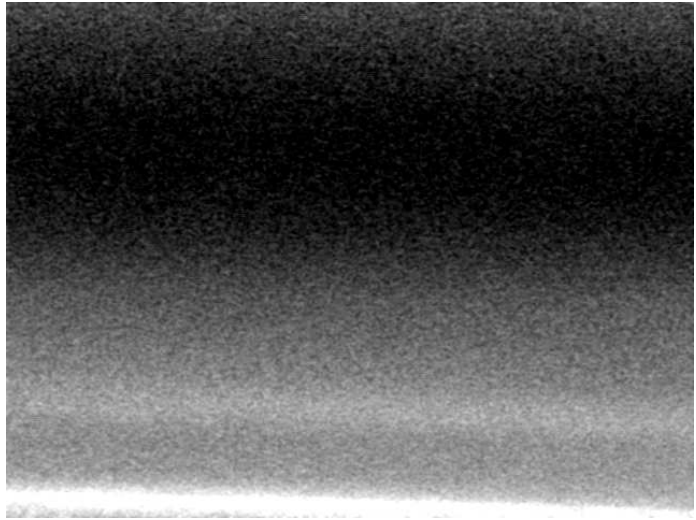}
\caption{\label{fig:emission1} The side-view obtained at 0 ${}^{\circ}$C substracted from the side-view at 70${}^{\circ}$ C, showing that at higher electrode temperature the intensity decreases in a broad region higher up in the sheath, while it increases in narrow regions right in front of the electrode surface.}
\end{figure}

Optical emission profiles in the sheath were obtained with the side-view camera. To ensure that particles did not influence the profiles, they were taken in the absence of dust. Figure \ref{fig:emission1} shows the side-view obtained in an experiment with fixed DC bias at 0${}^{\circ}$ C substracted from the side-view obtained at 70${}^{\circ}$ C. The images show the plasma glow fairly close to the lower electrode surface. The contrast has been enhanced in the image to increase visibility of the intensity differences. 

The dark region indicates that higher up in the sheath, the intensity is less at higher electrode temperature, while the thin bright structures show that at higher electrode temperature the local emission is increased in front of the electrode. These thin emission structures have recently been associated with the presence of fast neutral particles \cite{Donko2007} or energetic ions \cite{Nemschokmichal2008}. At the time these pictures were obtained, there was a slight mis-alignment between the camera and the lower electrode, which is the reason why the band structures are not precisely horizontal.

\begin{figure}[htbp]
\center
\includegraphics[width=1.0\textwidth]{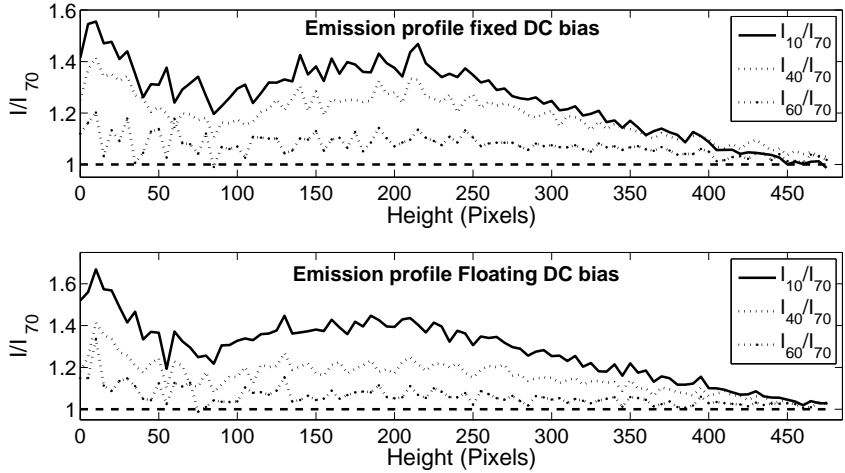}
\caption{\label{fig:emission2} Vertical emission profiles in the sheath normalized with respect to the emission profile at 70 ${}^{\circ}$C, for different electrode temperatures in the experiments with fixed DC bias (top) and with floating DC bias (bottom).}
\end{figure}

Figure \ref{fig:emission2} shows vertical emission profiles collected at 10, 40, and 60${}^{\circ}$ C divided by the one obtained at 70${}^{\circ}$ C. The total recorded height is 480 pixels, and the images are taken slightly higher up in the sheath than in figure \ref{fig:emission1}. The recorded hight corresponds to 12 mm, roughly the width of the sheath in front of the lower electrode. The profiles in both sets of experiments look similar, having a peak towards the bottom of the picture, corresponding to the highest horizontal band of figure \ref{fig:emission1}. For all temperatures the intensity becomes the same at the top of the image, which corresponds with the start of the plasma bulk, indicating that the bulk plasma emission is the same for all temperatures. For increasing electrode temperature, the observed intensity throughout a large part of the sheath decreases, however, indicating a change in the properties of the sheath.

\section{Discussion of results}

\subsection{Emission profiles}

Most of the visible line radiation in argon plasma at the powers and pressures discussed, is due to the de-excitation of neutral atoms in excited states rather than due to excited ionic states \cite{Palmero2007}, since the energy required for the direct excitation of ions is high (35 eV) and we do not expect a significant electron population with sufficient electron energy. The most likely excitation process is due to direct electron-impact excitation by energetic electrons (with $E>11.7$ eV), whereas the most likely de-excitation process is spontaneous de-excitation with the emission of a photon. There are three possible explanations for the observed reduction in emission: a reduction in the creation of excited atoms, a different de-excitation process (which does not involve the emission of photons in the visible spectrum) becoming dominant, or an increase in transport of the excited species away from their origin.

Due to the long life-time of meta-stable states, transport of such excited atoms away from their origin is possible. Since the GEC cell is constructed to minimize gas-flow, and since the temperature gradients induced by thermophoresis are too small to cause thermal convection \cite{Goedheer2009}, the excited neutral species will move through random Brownian motion. Thus, an increase in transport is possible with an increase in the neutral gas temperature. Since most energetic electrons are found in the (pre-)sheath, the main source of excited atoms is the (pre-)sheath region. Increased transport of these excited species might diffuse the emission coming from the (pre-)sheath region, explaining in part the reduced observed emission.

Different de-population mechanisms from excited states could include the excitation from such a state to a higher-energy state, which in turn spontaneously de-excites emitting a photon with a wavelength outside of the visual spectrum, or increased electron-impact ionization from the excited state. For the de-population to become important, a significant increase in electron density in the sheath would be required, since these processes require a second electron to collide with the excited atom within its life-time. Langmuir probe data in the sheath is not available, while details about the population of energy levels would require techniques such as line-ratio spectroscopy. We can therefore at this moment not exclude this possibility.

Finally, the easiest explanation would sipmly be a decrease in electron density and/or neutral gas density, since electron-impact excitation rates are proportional to the product of both densities. This possibility seems more likely, as is discussed below. A significant cooling of the electron population as a whole would also explain a reduction in the excitation processes, since an electron energy of at least 11.7 eV is required for the most dominant excitation transition. Again, we have no direct proof of this.

\subsection{Langmuir probe data}

\begin{figure}[htbp]
\center
\includegraphics[width=0.75\textwidth]{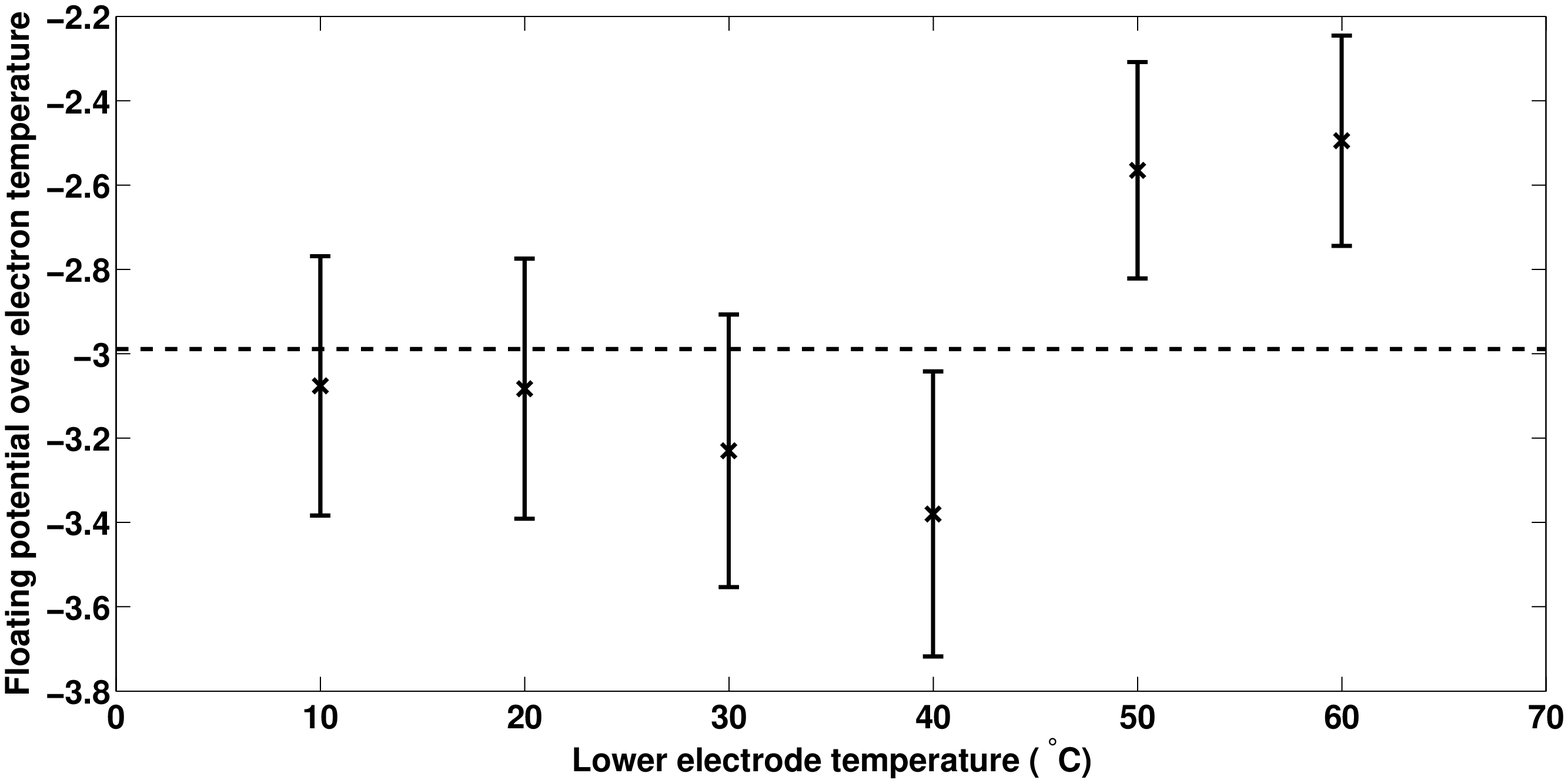}
\caption{\label{fig:ratio} $-K$ determined from Langmuir probe data of the floating potential, plasma potential and electron temperature; $-K = (V_{fl}-V_p)/T_e$. The measured values shown are very close to the value of -2.989 as determined in numerical simulations \cite{Goree1993}.}
\end{figure}

The Langmuir probe data presented above show no clear trend with temperature. According to numerical simulations, the local floating potential (which basically shows how negatively charged an object immersed in the local plasma would become) should be directly proportional to the electron temperature; $(V_{fl}-V_p) = -K T_e$ where $K$ is a constant that depends on the gas. At room-temperature in argon it was shown that $K=2.989$ \cite{Goree1993}. A plot of the ratio obtained from our data is shown in figure \ref{fig:ratio}, showing results consistent with theory. It also demonstrates a lack of a clear trend with electrode temperature. We therefore conclude that the bulk plasma parameters do not show a clear change with lower electrode temperature, which is consistent with the emission profiles.

\subsection{Electronic data}
In a low pressure RF discharge the sheaths act mostly capacitively, whereas the main power dissipation is due to the resistive bulk. In this case, the impedance of the plasma can be represented by two sheath reactances, $X_s$, in series with the bulk resistance $R_b$ \cite{Beneking1990}. Since the dissipative character of the sheaths is much smaller than the bulk, they are considered to be completely capacitive, and hence their impedance is purely imaginary. The discharge then has a complex \emph{impedance} $Z$ given by $Z=R_b+i(X_{sg}+X_{sp})$ where it is assumed that the grounded sheath and the powered sheath are not identical. 

\begin{figure}[htbp]
\center
\includegraphics[width=0.99\textwidth]{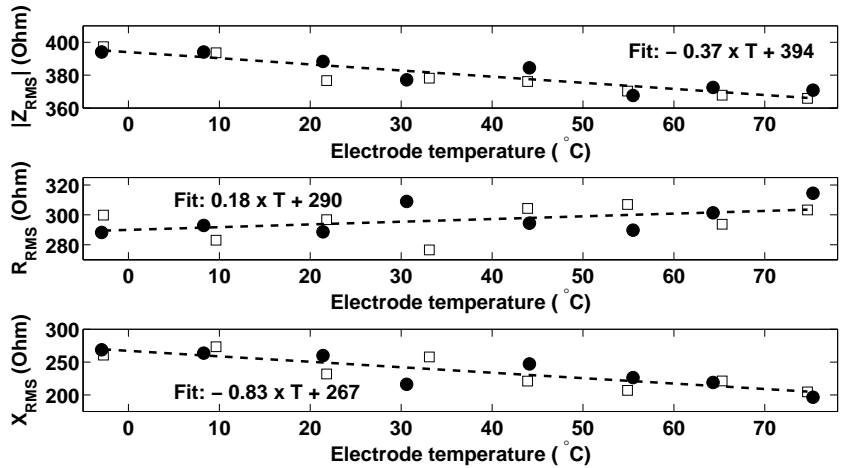}
\caption{\label{fig:E-data-results} The magnitude of the RMS impedance Z, the resistance R, and the reactance X of the plasma discharge against lower electrode temperature. The $\square$ corresponds to experiments with floating DC bias, the $\bullet$ to experiments with fixed DC bias. The dashed lines correspond to linear fits as noted in the graphs.}
\end{figure}

The absolute value of the impedance is given by $|Z| = V/I$, the ratio of the discharge voltage and current. The phase angle between the voltage and current, $\Phi$, is related to the resistance and reactance through $\tan(\Phi) = (X_{sg}+X_{sp})/R_b$, so that $R_b=|Z|/\sqrt{1+{\tan}^2(\Phi)}$. From our electronic data, we can thus calculate the RMS impedance, resistance and reactance. The results are shown in figure \ref{fig:E-data-results}. 

In a capacitively coupled RF discharge in argon at the pressures and powers considered, the bulk resistance can be approximated by $R_b = {\nu}_c/{\omega}_p^2{\epsilon}_0$, with ${\nu}_c = n_{gas}\int \sigma v_e f_e dv_e$ the electron-neutral collision frequency, $\sigma$ the cross section for elastic electron-neutral collisions \cite{Beneking1990}, which is approximately $3\times{10}^{-19}$ m${}^{2}$ for the electron energies in the discharge \cite{Raju2004}, and ${\omega_p} = \sqrt{n_ee^2/m_e{\epsilon}_0} $ the electron plasma oscillation frequency. Hence, $R_b \approx 3\times{10}^{-19} m_e n_{gas}\sqrt{8k_BT_e/\pi m_e}/n_e e^2 \approx {10}^{-7} (n_{gas}/n_e) \sqrt{T_e}$ Ohm, where we have assumed a Maxwellian distribution for the electrons. For a pressure of 200 mTorr, an electron density of about $5\times{10}^{14}$ m${}^{-3}$, and an electron temperature of 7 eV, we obtain $R_b \approx 370$ Ohm, which is very close to the determined values. 

The sheath reactance can be approximated by $X_s = d_s/{\omega}_{RF}{\epsilon}_0A \approx 1321 ~d_s/A$ Ohm, where $d_s$ is the sheath width and $A$ is the electrode surface area \cite{Beneking1990}. The sheath width can be approximated by the Child-Langmuir law, as $d_S \approx {10}^{-4} (\Delta V)^{3/4} J_{+}^{-1/2}$, with $\Delta V$ the voltage drop accros the sheath, and $J_{+}$ the ion current density across the sheath \cite{Godyak1993}. Assuming that the ion current density can be approximated by the Bohm flux $(J_{+} = e n_{+s}\sqrt{k_BT_e/m_{+}})$, and using the surface area of the lower electrode ($A = \pi (0.05)^2$), the sheath reactance becomes $X_s \approx {10}^{10} (\Delta V)^{3/4} n_{+} ^{-1/2} T_e^{-1/4}$ Ohm. For a potential jump across the sheath of 30 Volts, an ion density of $5\times {10}^{14}$ m${}^{-3}$ (assuming quasi-neutrality at the sheath edge), and an electron temperature of 7 eV, we find $X_s \approx 340$ Ohm, which is in good agreement with the determined values.

From the above, we see that a slight increase in resistance can result from an increase in gas density, a decrease in electron density, an increase in electron temperature, or a combination of these. On the other hand, the much stronger decrease in sheath reactance must result from a shrinking of the sheaths, which can result from an increase in plasma density, or electron temperature, or a smaller potential jump across the sheath. The observed change in plasma emission in the sheath might be an indication of these changes in the sheath width. Since the same decrease in reactance is also seen in the experiments with fixed DC bias, it is more likely that a change in the plasma parameters is responsible.

\subsection{Neutral atom density}

\begin{figure}[h]
\includegraphics[width=0.9\textwidth]{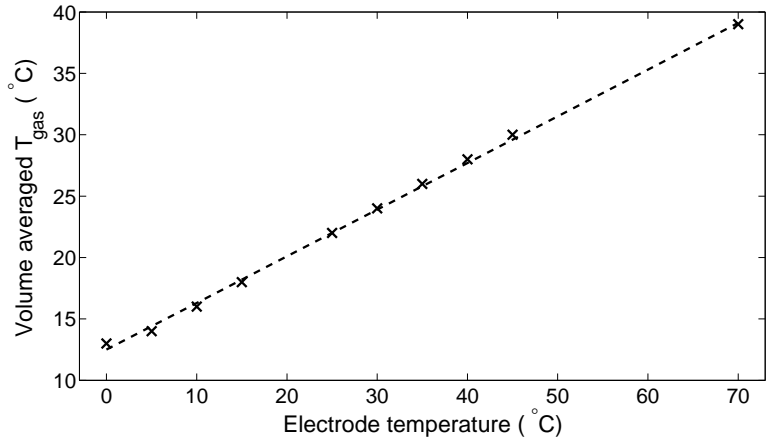}
\caption{\label{fig:gastemp} The volume averaged gas temperature in the discharge for various electrode temperatures, calculated using a self-consistent complex plasma fluid model. All other surfaces were kept at room-temperature (293 K, 20${}^{\circ}$ C). The temperature response can be well represented by a linear function, indicated by the dashed line.}
\end{figure}

Most complex plasma experiments, like the ones reported here, are performed at constant pressure. Heating or cooling the lower electrode causes heating or cooling of the gas inside the chamber, however. Even though one might suspect this to be a small effect, the change in temperature can be significant. Self-consistent fluid model calculations were done for our experimental setup for different electrode temperature settings. The volume averaged gas temperature computed with this model, which is used for self-consistent fluid simulations of complex plasmas \cite{Land2009}, is shown in figure \ref{fig:gastemp}. Assuming the ideal gas law, $P_{gas} = n_{gas}k_BT_{gas}$, we see that the density for 200 mTorr can vary from $6.6\times{10}^{21}$ m${}^{-3}$ at 0${}^{\circ}$ C to $6.0\times{10}^{21}$ m${}^{-3}$ at 70${}^{\circ}$ C, a difference of 10 \%. Such a change in neutral atom density will have a definite effect on both the global particle and power balance in the discharge, in turn affecting the electron temperature and the electron density. 

\subsection{Additional considerations}

Since the coverplate on top of the lower stainless steel electrode is made of aluminum, it is possible that a contact potential might be responsible for the change in DC bias. We therefore repeated the experiment without the coverplate and found the same result. This also eliminated secondary electron emission as a cause, since aluminum has (2-4) times the secondary electron emission coefficient of steel \cite{Bohm1992}, which means that any significant change in the DC bias behaviour would have been observable. The experiments were also repeated without the presence of particles, again showing the same results, indicating that the particles were not creating the observed change in DC bias.

One possibility we can not exclude is a change in surface-recombination chemistry, due to the increased electrode temperature \cite{Rebrov2003}. It is not unlikely that the surface accomodation might change with increased temperature, and this change might in part be responsible both for a change in the sheath behavior, as well as a change in the plasma parameters. At this time, no observational techniques are available in the CASPER laboratory to study this further. However, even if this were the responsible mechanism, it would still show that using thermophoresis in complex plasma changes the plasma parameters, indirectly coupling the thermophoretic effect to changes in the plasma parameters.

\section{Conclusion}

Using images of levitated dust particles, as well as of the plasma emission in the sheath, as well as electronic and Langmuir probe data, we have shown that thermophoresis has an effect on the discharge parameters in a capacitively coupled argon RF discharge. A slightly increased bulk resistance together with a more strongly reduced sheath reactance can be seen from the electronic data. Optical emission data seems to be consistent with this, in the sense that a clear change in the emission detected from the sheath region close to the electrode surface is observed, whereas the emission towards the plasma bulk does not change much with temperature. Langmuir probe data in the bulk is less conclusive, but seems to indicate a change in electron density, whereas the floating potential and electron temperature remain consistent with numerical estimates and do not show a clear trend with changes in the electrode temperature.

Numerical modeling indicates that experiments running at constant pressure have a varying neutral atom density when thermophoresis is applied, due to an increase in the volume-averaged gas temperature. This is expected to have a noticable effect on the plasma parameters, consistent with our observations. Therefore, reports on thermophoresis experiments should clearly indicate if the DC bias was allowed to float, or was fixed, since the bias has an important effect on the dust levitation height and both types of experiments give different results for the equilibrium levitation height. 

In the past, such information was considered unimportant. Our data seems to suggest that possibly some aspects of experiments in complex plasma involving thermophoresis might have been overlooked. In some recent experiments, thermophoresis played an important role, but the effects discussed in this report were not explicitly considered. Discussions in these works involved, for instance, the correct formulation of the thermophoretic force using dust particle levitation \cite{Rothermel2002}, the exact shape of the external confinement of particles levitated in a glass box with the help of thermophoresis \cite{Arp2005}, as well as observations of oscillations in a single layer of dust particles at an unusually thin edge (due to additional thermophoresis) of a void, even though in that particular case the temperature gradient arose naturally \cite{Liu2009}. Even though the relative changes in plasma parameters due to thermophoresis depend on the exact discharge settings, our results indicate that any explanation that assumes constant plasma parameters at different temperatures used to induce thermophoresis, are automatically invalid and can not describe the external confinement of dust particles in the discharge correctly for all temperatures. This is something that experimentalists in the field should be aware of.

\section*{Acknowledgements}
This work was supported by NSF under Grant PHY-0648869 and CAREER Grant PHY-0847127. The authors appreciate the discussion with many CASPER members. First author is especially grateful too for the discussions with Professor Overzet at UT Dallas.

\end{document}